\documentclass[a4paper,fleqn,usenatbib]{mnras}

\usepackage{graphicx}	% Including figure files
\title[Evidence for a Supermassive Black Hole in IRAS 20100$-$4156]{High-velocity OH megamasers in IRAS 20100$-$4156: Evidence for a Supermassive Black Hole}

\author[L. Harvey-Smith et al.]{L. Harvey-Smith$^{1}$\thanks{E-mail: lisa.harvey-smith@csiro.au}  
J.~R. Allison,$^{1}$ 
J.~A. Green,$^{1,2}$
K.~W. Bannister,$^{1}$
A. Chippendale,$^{1}$
\newauthor  
P.~G. Edwards,$^{1}$ 
I. Heywood,$^{1,3}$
A.~W. Hotan,$^{1}$ 
E. Lenc,$^{4,5,1}$
J. Marvil,$^{1}$
D. McConnell,$^{1}$
\newauthor 
C.~P. Phillips,$^{1}$
R.~J. Sault,$^{6,1}$
P. Serra,$^{1}$
J. Stevens,$^{1}$
M. Voronkov$^{1}$ and
M. Whiting$^{1}$
 \\
$^{1}$CSIRO Astronomy \& Space Science, PO Box 76, Epping, NSW 1710, Australia.\\
$^{2}$The SKA Organisation, Jodrell Bank Observatory, Macclesfield, SK11 9DL, U.K.\\
$^{3}$Department of Physics and Electronics, Rhodes University, PO Box 94, Grahamstown, 6140, South Africa.\\
$^{4}$ARC Centre of Excellence for All-sky Astrophysics (CAASTRO). \\
$^{5}$Sydney Institute for Astronomy, School of Physics A28, University of Sydney, NSW 2006, Australia. \\
$^{6}$School of Physics, University of Melbourne, VIC, 3010, Australia.
}

\date{Accepted by MNRAS on 22 April 2016}

\pubyear{2016}

\begin{document}
\label{firstpage}
\pagerange{\pageref{firstpage}--\pageref{lastpage}}
\maketitle

\begin{abstract}
We report the discovery of new, high-velocity narrow-line components of the OH megamaser in IRAS 20100$-$4156. Results from the Australian Square Kilometre Array Pathfinder (ASKAP)'s Boolardy Engineering Test Array (BETA) and the Australia Telescope Compact Array (ATCA) provide two independent measurements of the OH megamaser spectrum. We found evidence for OH megamaser clumps at  $-$409 and $-$562 km~s$^{-1}$ (blue-shifted) from the systemic velocity of the galaxy, in addition to the lines previously known. The presence of such high velocities in the molecular emission from IRAS 20100$-$4156 could be explained by a $\sim$50~pc molecular ring enclosing a $\sim$3.8 billion solar mass black hole. We also discuss two alternatives, i.e. that the narrow-line masers are dynamically coupled to the wind driven by the active galactic nucleus or they are associated with two separate galactic nuclei. The comparison between the BETA and ATCA spectra provides another scientific verification of ASKAP's BETA. Our data, combined with previous measurements of the source enabled us to study the variability of the source over a twenty-six year period. The flux density of the brightest OH maser components has reduced by more than a factor of two between 1988 and 2015, whereas a secondary narrow-line component has more than doubled in the same time. Plans for high-resolution VLBI follow-up of this source are discussed, as are prospects for discovering new OH megamasers during the ASKAP early science program.
\end{abstract}


\section{Introduction}

In this paper, we describe the results of a centimetre-wave radio study of  IRAS 20100$-$4156, an ultra-luminous infrared galaxy (ULIRG) with a redshift $z$ = 0.129 \citep{1989Natur.337..625S}.  IRAS 20100$-$4156 hosts a very luminous megamaser (actually a `gigamaser') in the $^2\Pi_{3/2}$, J = $\mathstrut{\frac {3}{2} }$, F = 2 $-$ 2 (1667.3590 MHz) main line transition of the hydroxyl (OH) molecule. OH megamasers are thought to occur when radio continuum emission from the galaxy is amplified by OH molecules whose energy-level population has been inverted by a far-infrared radiation field generated by dust \citep[see][for an overview]{2012msa..book.....G}. Optical and infrared images \citep{1990A&A...231L..19M, 2002ApJS..138....1B} reveal this to be part of a merging pair or possibly a trio of galaxies accompanied by tidal tails and clumps of star-forming gas and dust. Both optical and infrared data indicate that the system is in a starburst phase \citep{1999ApJ...514..660T}, has a Low Ionisation Nuclear Emission Region (LINER)-like spectrum \citep{1989Natur.337..625S} and a star formation rate, estimated from far-infrared emission, of approximately 700 solar masses per year \citep{2011ApJ...730...56W}. Searches for X-rays and polarised H$\alpha$ emission revealed no definitive evidence of an active galactic nucleus (AGN) \citep{2003MmSAI..74..442B,2003MNRAS.338L..13P}, however a \emph{Spitzer} Space Telescope study using a diagnostic based upon the 5 to 8~$\mu$m brightness ratio found evidence for a 20 per cent bolometric contribution from an AGN \citep{2010MNRAS.405.2505N}. This AGN is presumably deeply embedded within a dusty starburst region at the centre of IRAS 20100$-$4156.

The OH megamaser in IRAS 20100$-$4156 was discovered using the Parkes radio telescope by \cite{1989Natur.337..625S} and a slightly improved spectrum was later published by \cite{1992MNRAS.258..725S}. The former paper identified a double peak at 38,697 and 38,648 km~s$^{-1}$ in the barycentric rest frame with an OH peak flux density of 200~mJy and a broad `pedestal' of emission. IRAS 20100$-$4156 was subsequently observed using the Australia Telescope Compact Array (ATCA) as part of a study of magnetic fields in OH megamaser galaxies published by \cite{1996MNRAS.280.1143K}. They observed the line as having slightly lower peak flux density ($\sim$165 mJy) and found a previously unidentified (73 mJy) peak at approximately 38,500 km~s$^{-1}$, although an absolute flux calibration was made difficult by the lack of a robust understanding of the spectral baseline due to a narrow observing bandwidth. In this study, we provide a pair of independent observations of the OH megamaser in IRAS 20100$-$4156 in a third observing epoch. The observations were taken as part of spectral line commissioning of the Australian Square Kilometre Array Pathfinder (ASKAP) using the Boolardy Engineering Test Array \citep[hereafter BETA;][]{2014PASA...31...41H} and a follow-up observation was made using the ATCA. The aims of the study were to (i) test the spectral line capabilities of BETA, (ii) compare the results to those obtained with the ATCA and (iii) investigate the generation mechanism, variability and kinematics of the OH megamaser emission in IRAS 20100$-$4156.

\section[]{Observations}

BETA comprises six 12-metre diameter parabolic reflectors outfitted with prototype ASKAP phased array feeds \citep{schinckel11} based on a connected-element ``chequerboard" array \citep{hay08}. We used five antennas of BETA; antennas 1 (Diggidumble), 6 (Biyarli), 8 (Bimba), 9 (Gagurla) and 15 (Minda), to observe the galaxy IRAS 20100$-$4156 (Right Ascension 20$^{\mathrm{h}}$13$^{\mathrm{m}}$29$\fs$509, Declination -41$\degr$47$\arcmin$35$\farcs$48, J2000) from UT 20:00 March 3rd $-$ 08:00 March 4th 2015 (Scheduling Block ID 1496) and from UT 06:25 $-$ 10:42 on 12th May 2015 (Scheduling Block ID 1791). We also observed the radio galaxy PKS B1934-638 for 15 minutes on each day to provide calibration data before the target was observed. The correlator was configured to record a total bandwidth of 304 MHz between 1399.5 and 1703.5~MHz in 16,416 spectral channels, each with a width of approximately 18.5~kHz ($\Delta$$v$ = 3.76 km~s$^{-1}$ at 1477~MHz). Nine beams were generated, oriented in a 3 $\times$ 3 square pattern with an inter-beam spacing of 0.66 degrees, with the central beam pointed directly at the target source.

We subsequently obtained Director's discretionary time on the ATCA to provide an independent verification of our results. The observations (observing code CX322) were conducted in 6~km array configuration (6A) between UT 23:45 16th April $-$ 00:55 17th April 2015. The correlator \citep{2011MNRAS.416..832W} was configured to simultaneously record a 2 GHz-wide continuum band with 32 $\times$ 64 MHz channels (velocity resolution 11507.251 km~s$^{-1}$) and four overlapping zoom modes, centred at 1476 MHz. The output spectral line data set had 5120 spectral channels with a spectral resolution of about 31.3 kHz (or 5.619 km~s$^{-1}$). We again used PKS B1934-638 as a primary calibrator source. Observations of the target source (25 minutes) were interspersed with three short (2 minute) observations of the phase calibrator PKS 1954-388. The total integration time on the target source was 56 minutes, although a small amount of data was flagged due to a suspected phase jump on antenna 3.

\section[]{Data Reduction}

\subsection{BETA data reduction}

Our BETA data were flagged, imaged and calibrated following the method described by \cite{2015MNRAS.453.1249A}, although our procedure here deviates slightly. Notable changes include an increase in the spectral resolution of the full-band coarse-channel data to 1\,MHz, used for self-calibration. This was done to reduce the effect of radio frequency interference (RFI), which is more prevalent in this frequency band than at frequencies between 700~MHz and 1\,GHz. The flux scale was corrected using a model of PKS\,B1934-638 \citep{Reynolds94}, and we applied a primary beam correction assuming a simple Gaussian model, scaled with frequency. The BETA flux scale was verified against the Sydney University Molonglo Sky Survey source catalogue \citep{2003MNRAS.342.1117M} by \citep{2015MNRAS.452.2680S} and against the NRAO VLA Sky Survey catalogue \citep{1998AJ....115.1693C} by \cite{2016arXiv160105857H}. Since our target source is at the centre of the beam, we expect any systematic deviation in the flux density over the period of our observation to be dominated by variation in the gain from our observation of PKS\,B1934-638. We estimate the uncertainty in the BETA flux scale to be approximately 7\%. This was determined by adding in quadrature the 3\% uncertainty introduced by bootstrapping the flux to PKS\,B1934-638; a stochastic component of 4\% determined by comparing the flux densities of several continuum point-sources close to the field centre over the two BETA observing epochs; and an approximately 5\% component introduced by fitting and removing a baseline ripple introduced by nearby RFI. In Fig. \ref{Figure1} we show the resultant spectrum at the observed frequency of the red-shifted 18-cm OH line.

\subsection{ATCA data reduction}

The ATCA zoom mode (31.3 kHz spectral resolution) data were processed as follows. The raw uvdata were loaded into {\sc miriad} \citep{1995ASPC...77..433S} and some edge channels and those known to contain bad data or correlator artefacts were flagged using {\sc atlod}. Visibilities with extreme amplitudes ($>$500 Jy) were flagged using {\sc uvflag} and further interactive flagging was carried out using {\sc blflag}. Bandpass and polarisation leakage corrections were calculated for the primary flux calibrator PKS B1934-638 using {\sc mfcal} and {\sc gpcal}, respectively. Bandpass corrections were copied to the phase calibrator PKS 1954-388 before polarisation leakage corrections were determined using that source and the {\sc gpcal} task. Using {\sc gpboot}, the absolute flux scale of the phase calibrator was tied to PKS B1934-638, using the same method as for the BETA observations. The uncertainty in the absolute flux scale using this model was estimated by \cite{Reynolds94} to be 1$-$2\%. Adding this in quadrature with the 3\% systematic uncertainty introduced by bootstrapping the flux to PKS\,B1934-638 gives an overall uncertainty in the ATCA flux density scale of 4\%. We then adjusted the labelling of velocity channels in the target source data for the doppler correction during the observation using {\sc uvaver} and we converted the data to the barycentric rest frame. The calibration solutions from PKS 1954-388 were copied to the target source IRAS 20100$-$4156. A baseline ripple was reduced by fitting a fifth-order polynomial to the line-free channels surrounding the OH maser and subtracting this using {\sc uvlin}. The task {\sc invert} was used to generate a calibrated image cube of IRAS 20100$-$4156. The peak pixel in the cube was located using {\sc maxfit} and an average spectrum was extracted at that peak position using {\sc uvspec}. 

\section{Results}

\subsection{BETA results}

\begin{figure}
\includegraphics[width=8cm, trim=0cm 0cm 0cm 0cm, clip=true]{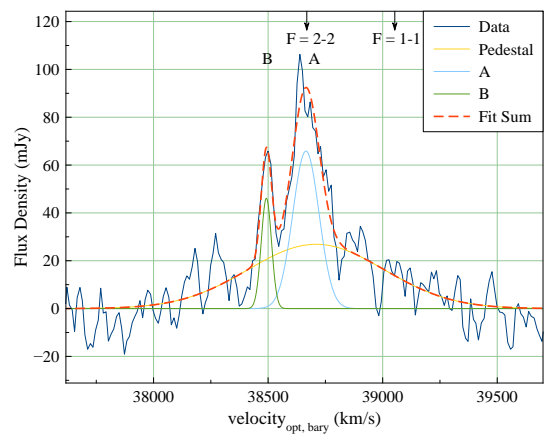}
\caption{Smoothed BETA spectrum of the OH megamaser in IRAS 20100$-$4156, showing the three Gaussian fits and their sum superimposed on the measured signal from BETA. We have labeled the narrow-line peaks A and B for reference. Peak A has a double structure, however the level of noise in the BETA spectrum means that our Gaussian fitting method does not recognise the double structure as a significant.}
\label{Figure1}
\end{figure}

The BETA spectrum shows significant spectral line emission in IRAS 20100$-$4156 lying close to the redshift-corrected rest frequency of the F=2$-$2 OH (1667~MHz) molecular transition. The median 1$\sigma$ root mean square (rms) noise in the BETA spectrum (determined using the full 304~MHz band) is 16.4~mJy per beam per spectral channel. Fig. \ref{Figure1} shows the observed spectrum, smoothed using third order Hanning smoothing. A simple baseline subtraction was carried out, however the left-hand side of the spectrum is still slightly affected by a baseline ripple caused by strong broadband radio emission from a satellite at approximately 1485~MHz. The main peak (A) apparently has a double structure, however the level of noise in the BETA spectrum means that our Gaussian fitting method recognises only a single component. 

We modelled the spectrum as the sum of three Gaussian components using {\sc magicplot}\footnote{http://magicplot.com/wiki/magicplot\_wiki\_home}. The model was determined by increasing the number of Gaussians until the residuals within the frequency range of the spectral line resembled the noise. {\sc magicplot} employs a Levenberg-Marquardt nonlinear least squares curve fitting algorithm to minimise the residual sum of squares ($\chi^{2}$) by iteratively varying the parameters of the input functions. Table 1 lists the characteristic velocities, widths and amplitudes of the three Gaussian components determined by the fitting routine and Fig. \ref{Figure1} shows the derived Gaussian fit curves and their sum, plotted with the measured data.

Our modelling reveals a broad component of OH emission, superimposed with two narrower peaks at approximately 38,665 and 38,492 km~s$^{-1}$ in the barycentric rest frame. The physical interpretation of the line structure is discussed in Section \ref{discussion}. We find no statistically significant emission at the redshift-corrected rest frequency of the OH F=1$-$1 (1665-MHz) line. The F=2$-$1 (1720-MHz) OH satellite line is a non-detection, which is unsurprising since the OH satellite lines are very rarely seen in megamaser galaxies \citep{2013ApJ...774...35M}. We were not able to observe the F=1$-$2 (1612-MHz) OH satellite line in these observations, as the redshifted rest frequency was outside our observed frequency range.

The continuum peak flux density of the source is $S_{\rm cont}$ =14.4$\pm$1.8~mJy, which was determined from an image made from the full 304 MHz frequency band centred at 1551.5~MHz observed with BETA. This is consistent with published measurements of $S_{\rm cont}$ = 21.1$\pm$1.4~mJy at 843 MHz from the Sydney University Molonglo Sky Survey \citep{2003MNRAS.342.1117M}, $S_{\rm cont}$ = 17.6$\pm$0.2~mJy at 1.49 GHz \citep{1996ApJS..103...81C} and $S_{\rm cont}$ = 9.6$\pm$3~mJy at 5 GHz \citep{1992MNRAS.258..725S} using Parkes, assuming a spectral index of $\alpha$ = $-$0.5, which is a typical value for ULIRGs at these frequencies \citep{2008A&A...477...95C}.

\begin{table*}
\centering
\caption{Measurements of the flux density, radial velocity and full-width at half maximum of the spectral peaks in the OH megamaser IRAS 20100$-$4156 from BETA data.}
\label{table1}
\begin{tabular}{|c|c|c|c|c|c|c|}
\hline
Peak ID & Flux density, $S_{\rm peak}$  & Peak velocity, v$_{\rm opt}$  & Gaussian FWHM \\
              & (mJy)  &   ($\times$10$^{3}$ km~s$^{-1}$) & (km~s$^{-1}$)  \\
\hline
Pedestal & 27 $\pm$17 & 38.71 $\pm$ 0.13 & 789 $\pm$ 326 \\
A & 66 $\pm$ 24 & 38.67 $\pm$ 0.02 & 135 $\pm$ 60 \\	
B & 46 $\pm$ 32 & 38.49 $\pm$ 0.02 & 53 $\pm$ 44  \\
\hline
\end{tabular}
\end{table*}

\subsection{ATCA results} 

\begin{figure}
\centering 
\includegraphics[width=8cm, trim=0cm 0cm 0cm 0cm, clip=true]{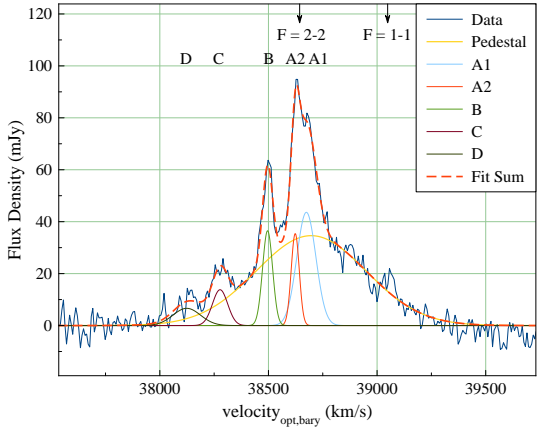}
\caption{Spectrum of the OH megamaser in IRAS 20100$-$4156 from the ATCA, showing Gaussian fits and their sum superimposed on the measured signal. Arrows show the positions of the redshift-corrected rest frequencies of the OH F=1$-$1 (1665-MHz) and F=2$-$2 (1667-MHz) lines. This spectrum shows the unsmoothed data, which has an rms noise of 3.9~mJy~beam$^{-1}$~channel$^{-1}$.}
\label{Figure2}
\end{figure}

Our ATCA spectrum, shown in Fig. \ref{Figure2} has a root mean-squared noise level of 3.9 mJy per beam per spectral channel. Six Gaussian components were fitted to the spectrum using {\sc magicplot}, yielding the peak flux densities, Gaussian full-widths at half maximum (FWHM) and peak velocities listed in Table \ref{table2}. We note that two further peaks (on the red-shifted side of the spectrum) are possibly visible. One of these lies very close to the redshifted frequency of the OH 1665~MHz line, which was also described as `tentatively' detected by \cite{1989Natur.337..625S}. After plotting the residuals we consider the significance of both of these features above the noise to be insufficient to include in our analysis.

The broad OH 1667~MHz pedestal and peaks A and B are consistent with those found with BETA. We also identified two additional spectral emission features in the ATCA spectrum, C and D, which lie several hundreds of km~s$^{-1}$ towards the blue-shifted side of the spectrum. These high-velocity components have not previously been reported in the literature and we discuss their physical interpretation in Section 5.3. We consider component D to be somewhat tentative, since its significance is sensitive to uncertainties in the width and position of the pedestal component. We include it here because it has significant flux lying wholly above the pedestal component across several spectral channels.

\begin{table*}
\centering
\caption{Measurements of the flux density, velocity and full-width at half maximum of the spectral peaks in the OH megamaser IRAS 20100$-$4156 from ATCA data.}
\label{table2}
\begin{tabular}{|c|c|c|c|}
\hline
 Peak ID & Flux Density, $S_{\rm peak}$ & Peak velocity, v$_{\rm opt}$ & Gaussian FWHM \\
               & (mJy)              & ($\times$10$^{3}$ km~s$^{-1}$) & (km~s$^{-1}$)   \\
 \hline
Pedestal & 34.5 $\pm$ 1.4 &  38.70 $\pm$ 0.01 & 600 $\pm$ 19  \\
A1 & 45.6 $\pm$ 2.5 & 38.67 $\pm$ 0.003  &	117 $\pm$ 6  \\
A2 & 26.5 $\pm$ 3.5 &  38.63 $\pm$ 0.002	& 36 $\pm$ 6  \\
B & 36.1 $\pm$ 2.1	&  38.50 $\pm$ 0.001	&  54 $\pm$ 4 \\
C & 13.7 $\pm$ 1.7 &  38.28 $\pm$ 0.01	& 82 $\pm$ 14 \\
D & 6.6 $\pm$ 1.3 & 38.12 $\pm$ 0.02  & 146 $\pm$ 42  \\	
\hline
\end{tabular}
\end{table*}

\section{Discussion}
\label{discussion}

\subsection{ATCA results verify BETA spectral line mode}

\begin{figure} 
\centering 
\includegraphics[width=8cm, trim=0cm 0cm 0cm 0cm, clip=truewidth]{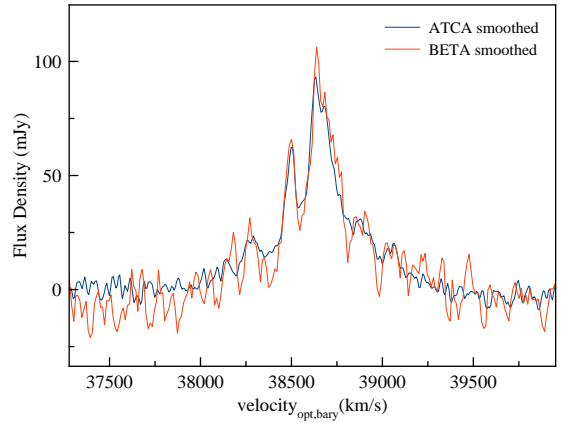}
\caption{Comparison of the spectra of the OH megamaser in IRAS 20100$-$4156 observed with BETA and ATCA. Both spectra have been smoothed with a 3-point central moving average. The BETA baseline on the left-hand side of the plot is affected by ripple from a satellite signal at 1485-MHz, which was present for part of our observation.}
\label{Figure3}
\end{figure}

One of the motivations for this work was to verify the spectral line performance of BETA. Fig. \ref{Figure3} shows the BETA and ATCA spectra of IRAS 20100$-$4156 plotted together. The spectral emission profiles are very similar, except for the fact that the blue wing of the BETA spectrum is impacted by a baseline ripple introduced by narrow-band RFI from a satellite that was broadcasting at 1485-MHz during our observations. Nonetheless, the positions and peak flux densities of the common spectral features agree within the measurement uncertainties (which are dominated by the spectral noise), confirming that the scientific results from the BETA spectral line observing mode are consistent with similar independent measurements.

The spectrum we obtained using the ATCA had a lower baseline noise level and has revealed a number of velocity components that have not previously been seen or reported in the literature. In the following section, we discuss the velocity structure and physical origins of these additional components of OH emission.

\subsection{Physical interpretation of the maser region}

Very luminous OH `gigamasers' occur in the most powerful ULIRGs and tend to have very broad pedestals and several strong narrow-line components \citep{2002AJ....124..100D}. This suggests that either the OH gigamasers originate in complex dynamical environments, or the central masses of their molecular rings are very high. Considering the generation mechanism for the maser emission, we now discuss these options for the origin of the OH gigamasers in IRAS 20100$-$4156. The model of \cite{2008ApJ...677..985L} describes OH maser emission generated in compact ($\sim$1pc) molecular clumps with turbulent line widths of approximately 20 km~s$^{-1}$, pumped by far-infrared radiation. In this scenario, the ensemble of maser clumps across the starburst region (typically the inner 100 pc of the galaxy in most starburst galaxies) produces a broad OH pedestal, whereas the brighter, high-gain masers occur when compact maser clumps happen to overlap in velocity along the line-of-sight. As noted by \cite{2001A&A...377..413P} and \cite{2005A&A...443..383P} these overlaps would occur most frequently at the tangents to a circumnuclear ring. Our observations with BETA and ATCA did not have sufficient angular resolution (46$"$ and 7$"$ respectively) to resolve the positions of individual maser spots. However, high-resolution VLBI studies have shown that narrow-line OH megamaser emission comes from the inner $<$50~pc of galaxies, although in some cases this is from two separate galactic nuclei within a merging system \citep[e.g.][]{1998ApJ...493L..13L, 1999ApJ...511..178D, 2001A&A...377..413P}.

The measured properties of the OH masers in IRAS 20100$-$4156 are compatible with a circumnuclear region origin. Our results reveal a broad OH pedestal spanning many hundreds of km~s$^{-1}$ punctuated by narrower peaks at +20, $-$20, $-$190, $-$409 and $-$562 km~s$^{-1}$ from the systemic velocity (v$_{sys}$ = 38,703~km~s$^{-1}$, determined by \cite{1989Natur.337..625S}). Each of these narrow-line peaks have velocity widths of a few tens of km~s$^{-1}$. This picture is similar to that presented by  \cite{2001A&A...377..413P} and \cite{2005A&A...443..383P}. We find that the flux density ratio of the pedestal to the background radio continuum emission is $S_{1667}$/$S_{\rm cont}$ = 0.54, suggesting that the diffuse maser has a low gain and is unsaturated, which is consistent with the \cite{2008ApJ...677..985L} model. In previous VLBI studies of OH megamasers, the narrow-line emission has been shown to be confined to a molecular ring with a radius $<$50 pc. Assuming this is the case in IRAS 20100$-$4156, the $v$=562 km~s$^{-1}$ spread of velocities of the narrow maser lines (using $M=v^2 r/G$) indicates an enclosed mass of $M<3.8 \times 10^9 M_{\odot}$. This is consistent with the black hole mass vs. H-band luminosity relation \citep{1997A&AS..124..533D, 2015ApJ...803...61M}. Further investigation using VLBI is warranted, since measuring the mass of black holes using maser emission is a very powerful technique to investigate the relationship between major mergers, black hole growth and quenching of star formation by powerful outflows from AGN \citep{2003Natur.421..821K}.

An alternative explanation for the high velocity components in IRAS 20100$-$4156 is an AGN-driven outflow. \cite{2013ApJ...775..127S} identified a very broad OH absorption feature in IRAS 20100$-$4156 from measurements of the 79 and 119~$\mu$m OH doublets. They found a red-shifted P Cygni component in the OH 79~$\mu$m spectrum that may indicate the presence of a very powerful molecular outflow. From these data, \cite{2013ApJ...775..127S} inferred a maximum (terminal) outflow velocity of 1609$\pm$200 km~s$^{-1}$, which is higher than in most ULIRGs, but consistent with those that host AGN \citep{2007ApJ...663L..77T}. If the compact maser clumps are dynamically associated with the outflow caused by the AGN, this may provide a potentially very interesting case study of AGN formation. However, it is not clear how this interpretation would tie in with our current understanding of OH megamaser pumping.

These lines of enquiry cannot be conclusively resolved based upon the current data. A twin approach of (a) higher sensitivity observations of this source (e.g. with the full complement of 36 ASKAP dishes) to determine the significance of the weaker lines and (b) high-resolution imaging of the maser emission using the Australian Long Baseline Array \citep{2015PKAS...30..659E} will provide the spatial positions of the bright, narrow-line masers and enable us to distinguish between these scenarios.

\subsection{Variability found in the OH megamaser}

By comparing our results to all other published observations dating back to 1988 we have found evidence for long-term variability of the OH megamaser in IRAS 20100$-$4156.  Table~\ref{table3} shows the peak flux densities of the three major components of OH maser emission over the available measurement epochs.  Over the past 26 years the flux densities of the major OH peaks have changed by ($-$59$\pm$32)\% for Peak A1, ($-$36$\pm$32)\% for Peak A2 and (+83$\pm$37)\% for Peak B. The dominant feature has shifted from A2 to A1 and the flux density of peak B apparently increased between 1988 and 1991, remaining largely stable since then. 

\begin{table*}
\caption{Time variability of the main peaks of the OH megamaser in IRAS 20100$-$4156. Peak OH flux densities are taken from the spectra published in \citet{1992MNRAS.258..725S}, \citet{1996MNRAS.280.1143K} and this paper (here we quote true peak values, not fitted models). The 1991 ATCA flux densities are likely to be slightly reduced from the true values since there was insufficient recorded bandwidth to carry out an effective baseline subtraction. The uncertainty in the Parkes flux density scale is approximately 30\%.}
\label{table3}
\centering
\begin{tabular}{|c|c|c|c|c|c|}
\hline
Date of Observation & Telescope & Flux Density & Flux Density & Flux Density & 1$\sigma_{rms}$ noise \\
         & (max. baseline)  & A1 (mJy) & A2 (mJy) & B (mJy)& (mJy)  \\
\hline
September 1988 & Parkes & 200 &148  & 35 & 5 \\
April 1991 & ATCA (3km) & 165 & 145 & 72 & 4.1  \\
March-May 2015 & BETA & 86 & 106 & 65 &  16.4 \\
April 2015 & ATCA (6km) & 82 & 95 & 64 & 3.9 \\
\hline
\end{tabular}
\end{table*}

The flux density of the Parkes continuum source IRAS~20100$-$4156 was quoted as 9.6$\pm$3 mJy \citep{1992MNRAS.258..725S}. This very large (30\%) uncertainty in the Parkes flux scale could potentially account for the apparent drop in the OH flux density between the 1988 epoch and subsequent observations listed in Table \ref{table3}. Instrumental changes and differences in absolute flux scales notwithstanding, the changes in the OH line profile shape point to genuine variability in the narrow-line masers, which may be due to (a) interstellar scintillation or (b) substantive changes in the physical conditions of the maser environment. This could arise from stochastic changes in the maser gain path length caused by turbulence, or a relative movement of maser clumps relative to the line-of-sight. Alternatively it could be related to a genuine reduction in the background source of radio continuum photons.

Previous studies of OH megamasers have in some cases found their narrow-line emission to be variable. \cite{2002ApJ...569L..87D} found a dramatic change in some, but not all, spectral features from IRAS 21272+2514, which the authors argued were likely to be very compact maser regions (smaller than 2 pc) that were undergoing interstellar scintillation. A dedicated monitoring program carried out with ASKAP or MeerKAT \citep{2009IEEEP..97.1522J} would be able to to distinguish between these scenarios and disentangle systematic measurement and calibration effects for IRAS 20100$-$4156.  

\subsection{Future Prospects}

This and previous studies \citep[e.g.][]{2015MNRAS.453.1249A} have demonstrated that the Murchison Radio-astronomy Observatory has a relatively clean spectral environment, particularly in the frequency range from 700 MHz to 1 GHz. The maximum redshift probed by the Arecibo OH Megamaser Survey \citep{2002AJ....124..100D} was effectively set by the RFI environment at Arecibo, with an effective lower limit on the observing frequency of 1200 MHz. The spectral environment was also mentioned as a significant constraint in the Green Bank Telescope OH megamaser survey \citep{2012IAUS..287..345W}. As terrestrial RFI impacts radio observatory sites around the world, ASKAP's location and radio-quiet protection \citep{wilson13} provide the opportunity to search for OH megamasers in the largely unexplored frequency range below 1200~MHz corresponding to $0.39 < z < 1.38$. This work is very encouraging for upcoming searches for megamasers in the Deep Investigation of Neutral Gas Origins \citep[DINGO;][]{2009pra..confE..15M} and the First Large Absorption Line Survey of H I (FLASH; Sadler et al. 2016, \emph{in prep.}) with ASKAP.

\section{Conclusions}

We have observed the OH megamaser in IRAS 20100$-$4156 using BETA and the ATCA. The OH maser peak flux density is approximately 100~mJy~beam$^{-1}$ and the continuum flux density at 1.4 GHz (using BETA) is 14.4~mJy. We compared the peak flux density, peak velocity and FWHM of the four major spectral features, finding that the parameters agree within the uncertainties of our Gaussian model fitting. This gives us an independent verification of the scientific performance of ASKAP's first-generation phased array feeds.

Our ATCA spectrum reveals at least two narrow-line spectral features that have not previously been reported. These lie at $-$409 and $-$562 km~s$^{-1}$ to the blue-shifted side of the systemic velocity. There is also a possible pair of weak features on the red-shifted side of the spectrum that warrant further investigation, particularly since one of these lies very close to the redshift-corrected velocity of the OH F=$1-1$ (1665 MHz) transition. We considered whether the high velocity masers originate from a compact circumnuclear ring or from the AGN-driven molecular outflow in IRAS 20100$-$4156. High-resolution imaging will solve this puzzle, however we note that a 50~pc circumnuclear ring would suggest the presence of an enclosed mass of 3.8$\times10^9~M_{\odot}$.

We studied the variation of the OH spectrum between 1988 and 2015, finding that the narrow spectral features have varied significantly in that time. This may be caused by interstellar scintillation or by intrinsic changes in the maser environment. A dedicated monitoring program carried out with ASKAP or MeerKAT will be useful to monitor IRAS 20100$-$4156 and understand more about the variation in its OH maser emission. Sensitive high-resolution observations of these masers using VLBI will enable us to locate the compact maser spots, confirm the extent of the circumnuclear masers and better constrain the black hole mass.

\section*{Acknowledgements}

The Australian SKA Pathfinder is part of the Australia Telescope National Facility which is managed by CSIRO. Operation of ASKAP is funded by the Australian Government with support from the National Collaborative Research Infrastructure Strategy. Establishment of the Murchison Radio-astronomy Observatory was funded by the Australian Government and the Government of Western Australia. ASKAP uses advanced supercomputing resources at the Pawsey Supercomputing Centre. We acknowledge the Wajarri Yamatji people as the traditional owners of the Observatory site. Parts of this research were conducted by the Australian Research Council Centre of Excellence for All-sky Astrophysics (CAASTRO), through project number CE110001020. The authors thank the referee, Andrew Walsh, for his insightful comments on the manuscript.

\bsp

\label{lastpage}

\end{document}